\documentclass[useAMS,usegraphicx]{mn2e}
\usepackage{multirow}
\usepackage{url}
\usepackage{rotating}
\usepackage{listings}

\title[The INTEGRAL/IBIS AGN catalogue I] 
{\emph {The INTEGRAL/IBIS} AGN catalogue I: X-ray absorption properties versus optical classification}
\author[A. Malizia et al. ]
{A. Malizia,$^1$\thanks{E-mail address: \texttt{malizia@iasfbo.inaf.it}}, 
L. Bassani,$^1$  A. Bazzano,$^2$ A. J. Bird,$^3$ N. Masetti,$^1$  \newauthor F. Panessa,$^2$ J. B. Stephen,$^1$ P. Ubertini$^2$ \\
$^1$ INAF/IASF-Bologna, Via P. Gobetti 101, I-40129 Bologna, Italy \\
$^2$ INAF/IASF-Roma, Via Fosso del Cavaliere 100, I-00133, Roma, Italy \\
$^3$ School of Physics and Astronomy, University of Southampton,       SO17 1BJ, Southampton, UK 
}

\begin{document}         

\date{Accepted .... Received ...; in original form ...}

\pagerange{\pageref{firstpage}--\pageref{lastpage}} \pubyear{2010}
\maketitle 

\label{firstpage}

\begin{abstract}
In this work we present  the most comprehensive INTEGRAL AGN sample.
It lists 272 AGN for which we have secure optical identifications, precise optical spectroscopy and  measured redshift values plus X-ray
spectral information, i.e. 2-10 keV and 20-100 keV  fluxes plus  column density. 
Here  we mainly use this sample to study the absorption properties of active galaxies, to probe new AGN classes and to test the AGN unification scheme.
We find that half (48\%) of the sample is absorbed while the fraction of 
Compton thick AGN is small ($\sim$7\%). In line with our previous analysis,  we have however shown that when the bias towards heavily absorbed objects 
which are lost if weak and at large distance is  removed, as it is possible in the local Universe, the above fractions 
increase to become 80\% and 17\%. 
We also find that absorption is a function of source luminosity, which implies some
evolution in the obscuration properties of AGN. Few peculiar classes, so far poorly studied in the hard X-ray band, have been detected and studied for the first time 
such as 5 XBONG, 5 type 2 QSOs and 11 LINERs.
In terms of  optical classification, our sample contains  57\% of  type 1 and  43\% of type 2 AGN; this subdivision is 
similar to that found in X-rays if unabsorbed versus absorbed objects are considered, suggesting that the match between optical and X-ray classification is overall good.
Only a small percentage of sources  (12\%) does not fulfill the expectation of the unified theory as we find 22 type 1 AGN which are absorbed and 10 type 2 AGN 
which are unabsorbed.  Studying in depth these outliers we found that most of the absorbed type 1 AGN have X-ray spectra characterized by either complex  or warm/ionized absorption
more likely due to ionized gas located in an accretion disk wind or in the biconical structure associated to the central nucleus, therefore unrelated to the toroidal structure.
Among 10 type 2 AGN which resulted to be  unabsorbed,  at most  3-4\% is still eligible to be classified as a "true"  type 2 AGN.
\end{abstract}


\section{Introduction} 
Observations of Active Galactic Nuclei (AGN) have revealed that many of them are obscured by material (dust and gas) located  
within the inner tens of parsecs of the central engine (see e.g. Bianchi et al. 2012 for a review) and intercepted by
our line of sight. This obscuring material will influence both the AGN and our observations, and the study of its properties is fundamental
for an unbiased understanding of AGN physics.  For example obscuration by gas and dust  is a key ingredient of  the AGN Unified Scheme, which proposes   that type 2
and type 1 AGN   are intrinsically the same type of object viewed from different orientation angles (Antonucci 1993): when our line of sight  does not intercept  any absorbing  material the source 
is classified as a type 1 AGN while in the opposite case, where we see obscuring gas and dust, the source is defined as a type 2 AGN. 
The absorbing material which determines the two main flavours of active galaxies is most likely  a toroidal structure (doughnut like or clumpy) thought to be present in all AGN. \\
How many objects are absorbed or not and what is the distribution of absorption among all objects are important information for AGN studies.
Cosmic X-ray background (CXB) synthesis models, in the context of the AGN Unification theory and based on a combination of absorbed and unabsorbed AGN, 
have been quite successful in reproducing the overall 
broadband spectral shape of the observed background (Gilli et al. 2007), but they need accurate observational information on the absorption parameter. 
Absorbed sources  constitute also an important ingredient for the IR and the sub-mm backgrounds, where most of the absorbed radiation is re-emitted by dust 
(e.g. Fabian \& Iwasawa 1999).\\
Many important issues related to the population of absorbed AGN are still to be understood, like the number of type 2 QSO,  the nature of the 
X-ray bright optical normal galaxies (XBONG), the role of LINERs and the relationship between optical absorption and X-ray obscuration. \\
Because of the effect it may have on observations,  absorption can be problematic for surveys performed at various wavebands and this is
the main motivation for performing AGN  surveys above 20 keV.\\
In this work we have gathered a large number (272) of AGN detected by INTEGRAL, collected their optical classification  and X-ray column density measurements  in order to be able to 
study the absorption properties in a large  sample of hard X-ray selected AGN, explore the nature of absorption in XBONGs, type 2 QSO's and LINERs and finally  
compare the optical classification with the X-ray absorption as a tool to test the AGN Unification Theory.\\
It is worth noting that these studies have often been performed employing samples of AGN observed in
the soft (below 20 keV) X-ray band and  only recently with the advent of enough AGN being
detected at higher X-ray energies  have similar studies been attempted for the first time.\\
The paper is organized as follows: the INTEGRAL AGN catalogue is presented in section 2, the study of absorption properties including a discussion on the fraction of 
heavily absorbed objects is reported in section 3, peculiar objects belonging to the LINER, XBONG and type 2 QSO classes are presented in section 4.
In section 5 the optical classification versus X-ray absorption has been treated and in subsections 5.1 and 5.2 the absorbed type 1 AGN and unabsorbed type 2 AGN have been
discussed. The summary and conclusions are drawn in section  6.

\begin{figure}
\includegraphics[width=1.0\linewidth]{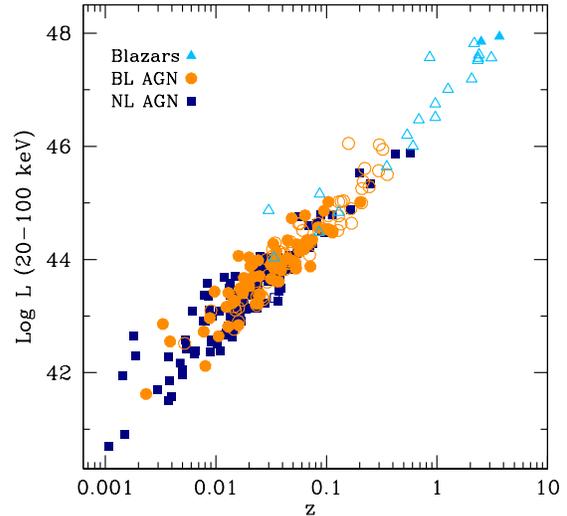}
\caption{Hard X-ray luminosity vs redshift for all the INTEGRAL AGN sample. Circles are broad line (BL) AGN, squares are narrow line (NL) AGN and triangles are blazars
(see section 5 for a more detailed classification). Filled symbols represent objects where  intrinsic absorption in excess to the Galactic one has been measured while open 
symbol refer to sources where there is only an upper limit to the column density, including Galactic values.}
\label{fig1}
\end{figure} 

\begin{figure}
\includegraphics[width=1.0\linewidth]{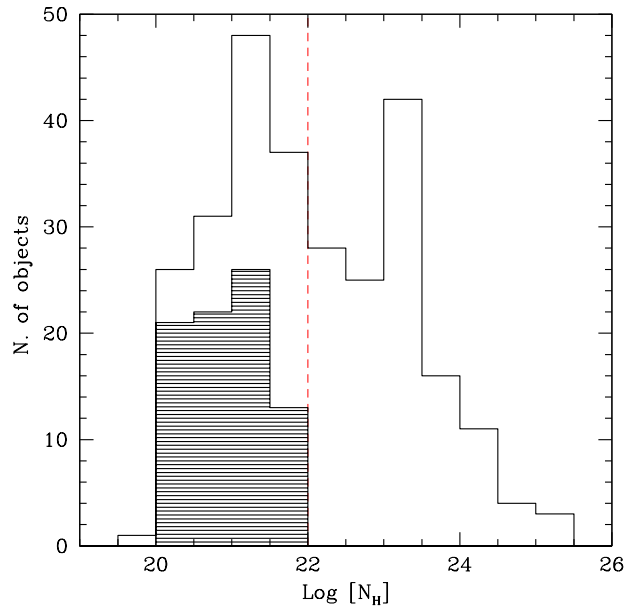}
\caption{Distribution of column densities in the INTEGRAL AGN. The dashed bins represent upper limit measurements including Galactic values.}
\label{fig1}
\end{figure}

\section{The INTEGRAL AGN catalogue}
In the 4th INTEGRAL/IBIS survey (Bird et al. 2010) there are 234  objects which have been identified with AGN. To this set of sources, we have then added 
38 galaxies listed in the INTEGRAL all-sky survey by Krivonos et al. (2007) updated on the website\footnote{http://hea.iki.rssi.ru/integral/survey/catalog.php}
but not included in the Bird et al. catalogue due to the  different sky coverage (see source names in bold in the Appendix). 
The final dataset presented and  discussed in this work
therefore comprises 272 AGN (last update March 2011), which represents the most complete view of the INTEGRAL extragalactic sky to date.
Although  new source identifications/classifications  are continuously coming  in (e.g. Masetti et al. 2012), they are not considered in the present dataset because they are not
properly characterized  in X-rays (see below).
In the Appendix we present the full catalogue, listing all 272 INTEGRAL AGN together with their optical coordinates, redshift, class, 20-100 keV flux  and X-ray data (2-10 keV flux, 
column density and reference work from which these data  have been taken). \\
The sample has  two great strengths: a) all sources have optical spectra, which means a secure identification and a measured redshift (the only exception is the BL LAC object RX J0137.7+5814 for which no redshift measure is available),
and b) all sources have X-ray data available, which provides a measure of the intrinsic absorption in each source.
Until recently, 34 sources did not have  X-ray (2-10 keV) coverage mainly due to the fact that they are newly discovered AGN; for these sources  X-ray data 
available from the Swift and XMM-Newton archives have been analysed and the results were published by Malizia et al. (2011) except for IGR J04221+4856 which has been recently observed by
Swift and the measurements of 2-10 keV flux and column density are reported here for the first time (see table in the Appendix).
It is also worth noting that for each source we have verified that the X-ray counterpart of the IBIS object corresponds to the optical identification.\\
In order to assign to each object the most appropriate optical class, we have searched the literature thoroughly and when possible  have also compared  the class reported 
in NED (NASA Extragalactic database) with that listed in the 13th edition of the  Veron-Cetty, Veron  extragalactic catalogue (2010); a significant number of AGN 
in the sample have been classified through our own follow up work (Masetti et al. 2012 and references therein). In these cases, we have adopted the classification 
criteria of Veilleux \& Osterbrock (1987) and the line ratio diagnostics of both Ho et al. (1993, 1997) and
Kauffmann et al. (2003);  for assigning  Seyfert subclasses (1.2, 1.5, 1.8, and 1.9), we have  used the H$\alpha$/[OIII]$\lambda$5007 
line flux ratio criterion described in Winkler (1992).
We have generally preferred the most recent classification and, in case of conflicting results, we have always checked  the optical spectra before assigning the most appropriate optical type.\\
The overall result is a list of AGN properly characterized at  optical, soft and hard X-ray frequencies thus available for population studies.\\
In figure 1 the whole sample is reported in the classical 20-100 keV luminosity $vs$ redshift plot.
The luminosities have been calculated for all sources assuming H$_{0}$=71 km s$^{-1}$ Mpc$^{-1}$ and q$_{0}$ = 0.  
All AGN have been plotted using three different symbols and following the optical classification (broad line (BL) and narrow (NL) AGN plus blazar); a more detailed analysis of
the various optical classes also in terms of absorption, can be found in section 5.
From figure 1 it can be estimated that our sensitivity limit is around 1.5 $\times$ 10$^{-11}$ erg cm$^{-2}$ s$^{-1}$. 
We find that the source redshifts span from 0.0014 to 3.7 with a mean at $z$=0.1477  and a peak in the distribution at $z$ = 0.015;
while the Log of 20-100 keV luminosities (in erg s$^{-1}$) ranges from  40.7 to $\sim$48 with a mean at around 46,  although the peak of the distribution is at  43.9.
NGC 4395 (a Seyfert 2)  is the closest and least luminous AGN seen by INTEGRAL, while IGR J22517+2218 (a broad line QSO) is the farthest and most luminous one;
the former hosts a relatively  small central black hole (M $\sim$10$^{4}$-10$^5$ M$_\odot$, Filippenko \& Ho 2003) while the latter houses a more massive object (M = 10$^9$ M$_\odot$, De Rosa 
private communication).
In conclusion, the present INTEGRAL sample spans a large range in source parameters and is therefore representative of the  population of AGN selected in the
hard X-ray band.

\begin{figure}
\includegraphics[width=1.0\linewidth]{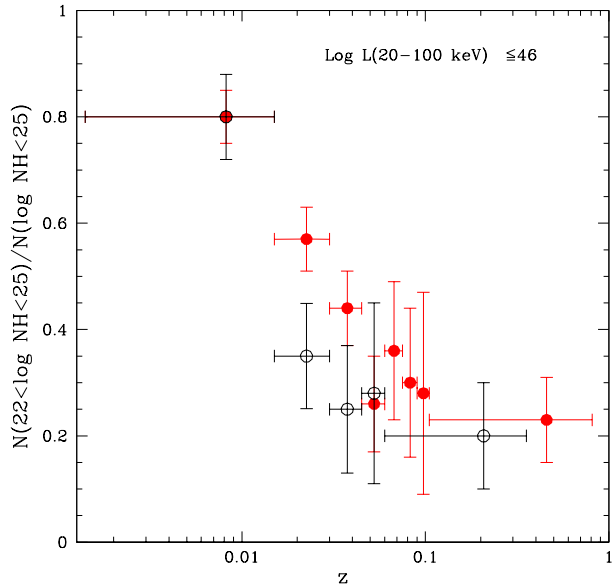}
\caption{Fraction of absorbed objects compared to the total number of AGN as a function of redshift in the present work (red points) and in the INTEGRAL complete sample of Malizia et al. 2009
(black points).}
\label{fig1}
\end{figure}

\begin{figure}
\includegraphics[width=1.0\linewidth]{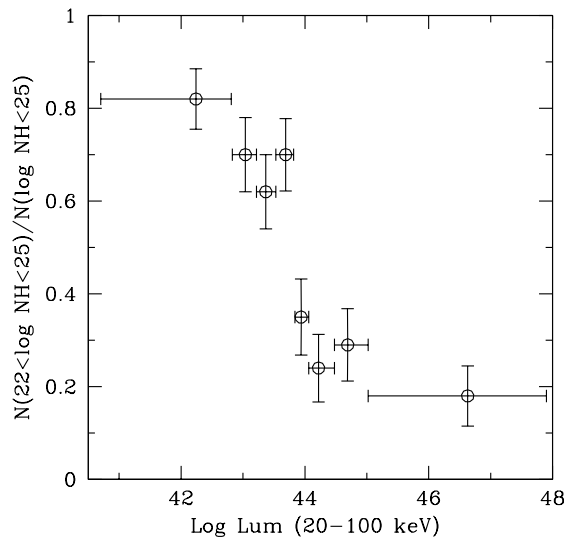}
\caption{Fraction of absorbed objects compared to the total number of AGN as a function of 20-100 keV luminosity.}
\label{fig1}
\end{figure}

\section{Absorption properties}
Since hard X-ray selected samples provide the most accurate estimate of the fraction of  absorbed objects as well as of Compton thick AGN, we can use the present sample for this purpose.
The column density distribution for the entire sample is shown in figure 2. 
Here and in the following we assume  N$_{H}$ = 10$^{22}$ cm$^{-2}$ as the dividing line between
absorbed and unabsorbed sources: this is the value conventionally used because it  corresponds to a column density sufficiently high  to hide the 
broad line region (BLR, Silverman et al. 2005). 
It is also worth noting that for a number of objects we did not measure any absorption in excess of the Galactic value and  have therefore used the Galactic column  
density as an  upper limit to the source intrinsic absorption.
With the assumptions made above the fraction of absorbed objects present in our sample is 48\% as can be seen in figure 2. \\
Within our catalogue we find 15 mildly (1.5 $\times$ 10$^{24}$  $\le$ N$_{H}$ $\le$ 10$^{25}$ cm$^{-2}$)  and 3 heavily (N$_{H}\ge$ 10$^{25}$ cm$^{-2}$) 
Compton thick AGN (the full list of objects can be extracted from the table in the Appendix ); we therefore estimate the fraction of Compton thick objects to be
around $\sim$7\%, in full agreement with estimates available in the literature (Malizia et al. 2009, Burlon et al. 2011).\\
Despite the fact that hard X-ray instruments are the least biased in terms of detecting absorbed AGN, they still miss some 
Compton thick objects, essentially those with weak (intrinsic) fluxes and at large distances. This has been fully discussed by Malizia et al. (2009)
and Burlon et al. (2011), who have shown  that once the correction for this bias is applied the real intrinsic fraction of Compton-thick AGN is around 20-24\%.
In particular, in our previous work (Malizia et al. 2009) we have adopted a redshift cut (z=0.015 or 60 Mpc) in a complete sample of INTEGRAL AGN in order to remove the bias and
to probe, although only locally, the entire  AGN population.
Following this reasoning and using the present much larger sample of objects,
we should  be able to expand this study  and to confirm our previous hypothesis, having in mind that the present sample is not complete.
For comparison with our previous study, we have  divided our sample in the same bins of redshift (up to z=0.57 considering only the AGN with Log L$_{20-100~keV} \leq$ 46) 
and plotted  the fraction of absorbed (N$_{H}>$10$^{22}$ cm$^{-2}$) objects compared to the total number of AGN in these bins.
The result is shown in figure 3 for the present sample (red points) and for that used in Malizia et al. (2009) (black points). \\
A number of considerations can be made from this figure:
first, the bias is still present  as we keep observing  a  trend of decreasing
fraction of absorbed objects as the redshift increases; second, we note that in the first bin the fraction of absorbed  objects remains the same as found in our previous analysis: 
in particular we find that, over the 66 objects present in this bin, 53 (or 80\%) have a column density $\ge$  10$^{22}$ and 11 (or 17\% $\pm$ 3\%) are Compton thick. 
Taking into consideration the fact that this is not a complete sample, these results are  in close agreement with those found previously and confirm 
our original suggestion that our survey is able to pick up all AGN, even the most absorbed, but only in the local Universe.
Finally, we note that the fraction of absorbed (and Compton thick) AGN has increased in the second bin from 35\% to 57\% (including 4 Compton thick sources), 
implying that as the INTEGRAL survey enlarges,  we are able to pick up more absorbed  objects  among those  which are  
distant and faint and therefore lost in previous catalogues.
We have also looked for a trend of decreasing fraction of absorbed AGN with increasing source hard X-ray luminosities. This effect, which is well documented in the X-ray band 
(Ueda et al. 2003, Hasinger et al. 2005, La Franca et al. 2005, Della Ceca et al. 2008),
 has also been observed at higher energies (Bassani et al. 2006, Sazonov et al. 2007 and references therein) and is found in our sample too.
In figure 4 we show how the fraction of the AGN with N$_{H}$ $\geq$ 10$^{22}$ cm$^{-2}$ in the INTEGRAL sample changes with 20-100 keV luminosity where
the width of each bin has been chosen so that the number of objects in each is constant ($\sim$34).
Again, the fraction of absorbed sources is around 80\% at low luminosities, decreasing to $\sim$20\% at higher luminosities.
At the moment it is not possible to discriminate as to whether the redshift effect  may have contaminated this result or if it is a direct consequence of the evolution of AGN 
luminosity function with $z$, but this issue will be addressed in more detail in a dedicated paper.\\
Summarizing our results, we conclude that the bias effecting deep hard X-ray surveys of AGN is real but negligible if we deal with objects located in the nearby Universe, 
where  the "true" fraction of absorbed  and Compton thick objects can be estimated with some precision at 80\% and $\sim$17\% respectively.

\section{Peculiar sources}
A few interesting classes of objects are now emerging in the INTEGRAL surveys; these   
are discussed in some detail in the following sub-sections.	

\subsection{XBONG}
One of the most interesting findings of recent soft X-ray surveys is the existence of XBONGs, i.e. X-ray Bright Optically Normal Galaxies (Comastri et al. 2002).
These sources are characterized by an X-ray luminosity of  10$^{43}$-10$^{44}$ erg s$^{-1}$ but are 
optically dull, i.e. they are hosted by normal galaxies whose optical spectra show no emission lines. 
Different interpretations have been suggested to explain these unusual properties (Trump et al. 2009)
including: heavy obscuration by gas covering
almost 4$\pi$ of the nuclear source; a Radiatively Inefficient Accretion Flow (RIAF)
which provides  a featureless hard X-ray spectrum and negligible emission in the optical and UV bands;  and finally dilution of nuclear emission from the host galaxy
starlight  which prevents the detection of the AGN optical spectrum (Moran et al. 2002).\\
Although  these explanations provide
a good description of the observed properties of a few objects, the nature of XBONGs is still the subject of
debate and recent studies suggest that they could be a mixed bag of different source typologies (Civano et al. 2007).\\
In the present sample of INTEGRAL AGN there are  5 objects classified as  XBONG (see table in the Appendix);  
all 5 are heavily absorbed in X-rays (Log N$_{H}$ from $\sim$23 up to $\ge$24 cm$^{-2}$) and quite bright 
(L$_{20-100~keV}$ in the range 6$\times$10$^{42}$-6$\times$10$^{44}$ erg s$^{-1}$). Clearly, option one is a viable explanation to account for their properties,
while it is difficult to reconcile a RIAF accretion model with the X-ray brightness of our objects.  We also note that 4 out of 5 sources are optically 
very bright as they are listed in the USNO B1 catalogue with R $\le$10.3; the only exception is IGR J17009+3559 which has R=14.4 but is  also the XBONG  with the highest redshift (z=0.13). 
This suggests that the dilution hypothesis (bright galaxy host) combined with heavy obscuration may be the explanation for the peculiar 
optical classification of our objects.

\subsection{Type 2 QSO}
Type 2 QSO are the high redshift/high luminosity  counterparts of local Seyfert 2 galaxies. They play an important role in our understanding of the Universe as their existence in considerable numbers is needed for the synthesis of the X-ray background (Gilli et al. 2007). 
They are also important objects in the subject of the evolution of  absorption with intrinsic luminosity
and/or redshift; this is an issue which  is still matter of intensive debate, but potentially very important for the influence that it can have on many astrophysical issues.\\
Despite much effort put into their search/quest, very few type 2 QSO have so far been found especially at very high redshifts. 
They are hard to discover because a significant fraction of their emitted power is absorbed by an optically thick torus  and, 
since they are distant/faint, they are missed by most surveys;  in particular, the 
lack of emission lines over a wide  wavelength range hampers their identification in the optical.
Hard X-ray surveys are ideally suited for this purpose as they can penetrate the torus in most objects.
The definition of type 2 QSO is somewhat arbitrary and depends on the waveband used to find them.
In X-rays the definition first introduced by Mainieri et al. (2002)  is generally adopted: a  type 2 QSO has to have an intrinsic X-ray luminosity
of L $>$ 10$^{44}$ erg s$^{-1}$ (0.5-10 keV) and an absorbing hydrogen column density of N$_{H}$ $>$ 10$^{22}$ cm$^{-2}$. 
Unfortunately,  we do not have 0.5-10 keV luminosity information for all our objects, 
but  can adopt the 2-10 keV band as a measure of the X-ray brightness and the same threshold in luminosity ($>$ 10$^{44}$ erg s$^{-1}$)
and column density (N$_{H}$ $>$10$^{22}$ cm$^{-2}$) used by Mainieri et al. (2002).\\ 
We have 12 sources fulfilling the above  criteria; however  4 of then  are local Seyferts ($z<$0.1) and cannot be considered as type 2 QSO
while 2 of them, PKS 1830-211 and IGR J22517+2218 are high redshift blazars 
in which the high column density measured may not be related to the toroidal structure implied by the AGN unified theory. 
PKS 1830-211 (z = 2.507) is a complicated system,  gravitationally lensed by an intervening galaxy at z = 0.89.
The observed flattening in the X-ray spectrum at low energies  has  been interpreted in other ways: either as absorption  coming from the lensing galaxy,  
or due to a change in the source spectral shape related to its blazar  nature (Zhang et al. 2008).
This last  interpretation has also been used in the case of the other absorbed blazar  IGR J22517+2218 (Bassani et al. 2007). 
The other source to explain is IGR J10147-6354, classified as a Seyfert 1.2 at z=0.2, 
 where the absorption, as well as in other type 1 AGN, is less easy to understand and will be the subject of an in-depth analysis in section 5.\\
We are left with 5 objects which display evident narrow emission lines in their optical spectra and therefore qualify to be type 2 QSO: 
IGR J00465-4005, SWIFT J0216.3+5128, IGR J09523-6231, IGR J12288+0052 and IGR J23524+5842.
Only IGR J09523-6231 is a type 1.9 AGN implying that  a weak broad H$\alpha$ emission line is present in its optical spectrum.
The  2-10 keV luminosity of these objects goes from 3.6 $\times$ 10$^{44}$ erg s$^{-1}$ to 5.6 $\times$ 10$^{45}$ erg s$^{-1}$ while  in the 20-100 keV band ranges  from
7.6 $\times$ 10$^{44}$ erg s$^{-1}$ to 7.8 $\times$ 10$^{45}$ erg s$^{-1}$.
The maximum absorption is measured in IGR J00465-4005 with N$_{H}$ $\sim$ 2.4 $\times$ 10$^{23}$ cm$^{-2}$ (Landi et al. 2010a).  
Despite early indications that IGR J12288+0052 could be a Compton thick AGN (Vignali et al. 2010) we find that it is 
only  moderately absorbed (Fiocchi et al. 2010, but see considerations in section 5.2); it is also the source at  the highest redshift (0.5756) within this group.\\
Type 2 QSO constitute  therefore a small fraction (2\%) of the entire AGN population selected in the hard X-ray band; they are 
outnumber by type 1 QSO implying that are either more rare or more difficult to find.\\ 
Finally, type 2 QSO have often been associated with very red objects  and some authors (Gandhi et al. 2004; Severgnini et al. 2005) have even suggested a  connection with ERO (Extremely Red Objects).
ERO are characterized as having R-K colour $>$ 5 (Severgnini et al. 2005 and references therein). 
We have collected this information for 4 out 5  sources and found that R-K  is in the range 2.3-5.6; only IGR J12288+0052
is marginally compatible with being a ERO but  its R-K value is only an upper limit. 
We can therefore conclude that  probably none of our sources qualifies to be an ERO .

\subsection{LINERs}
Low-ionization nuclear emission regions (LINERs) were identified as a class of galaxies by Heckman (1980) based on the relative intensities of their oxygen emission lines.
Suggestions for the ultimate power source of these objects include (1) a weak active galactic nucleus  harboring an accreting, supermassive black hole, (2) hot stars (either young or old)  
and (3) shocks. Recent radio, UV, and X-ray surveys at high spatial resolution, mid-IR spectroscopy, and variability studies have uncovered weak AGN in the majority of LINERs studied so far, suggesting that they make up a large (perhaps the largest) subset of all AGN  (Eracleous et al. 2010).  
Furthermore, as suggested by Ho (1999) the detection of broad lines in the optical spectra of many LINERs  constitute  strong evidence in favour of the AGN interpretation.  
As Seyfert galaxies, also LINERs can be classified in two classes, 1 and 2, depending on whether or not a broad component in their optical spectra can be seen.\\
In our sample,  a number of objects (11) have LINER features: in the optical diagnostic diagrams used for AGN optical classification,  
6 fall at the  boundary with Seyfert 2 galaxies and 3 at the boundary with Seyfert 1 galaxies;
the first  are absorbed in X-rays while the second are not.  Two objects lie in the region occupied  by "pure" LINERs and 
are also absorbed in X-rays; both display only narrow lines in their optical spectra. So we conclude that, like Seyfert 
galaxies, also LINERs come in two flavours: unabsorbed type 1 and absorbed type 2. 
Given their high X-ray luminosities, it is almost certain that all of these objects  are powered by AGN, although their 20-100 keV mean luminosity
($\sim$1.7 $\times$ 10$^{43}$ erg s$^{-1}$) is slightly lower than that of  Seyfert galaxies ($\sim$5.5 $\times$ 10$^{43}$ erg s$^{-1}$). 

\begin{table}
\begin{center}
\caption{Optical classification of sources in the INTEGRAL sample.}
\begin{tabular}{clr}
\hline
Class                     &  Spectral type   &    Number  \\
                              &                          &                    \\
 \hline\hline
Type 1                   &     Sy 1             &    46            \\
 (57\%)                   &     Sy1.2           &    24            \\
                              &     Sy1.5           &    45            \\
                              &     Sy1/LINER   &    3              \\
                              &     NLS1            &   14          \\
                              &     Blazar           &   22         \\
Type 2                   &     Sy 2             &   69           \\
  (43\%)                            &     Sy1.9           & 13           \\
                              &     Sy2/LINER   & 6     \\                           
                              &    Radio Gal.     &  1 \\                                                                                  
                            &     XBONG        &  4$^{\dagger}$  \\
                            &     LINERs         &  2 \\  
                            &     QSO 2           & 5 \\
                            &     Compton thick$^{\dagger}$ & 18  \\                                
\hline                     
Total                     &                          &     272 \\                           
\hline
\hline
\end{tabular}
\end{center}
$\dagger$ all the  Compton thick sources are Seyfert 2 but , NGC 1365, which is optically classified as 1.9 and MCG-07-06-018, which is optically classifies as XBONG.
\end{table}

\begin{figure}
\includegraphics[width=1.0\linewidth]{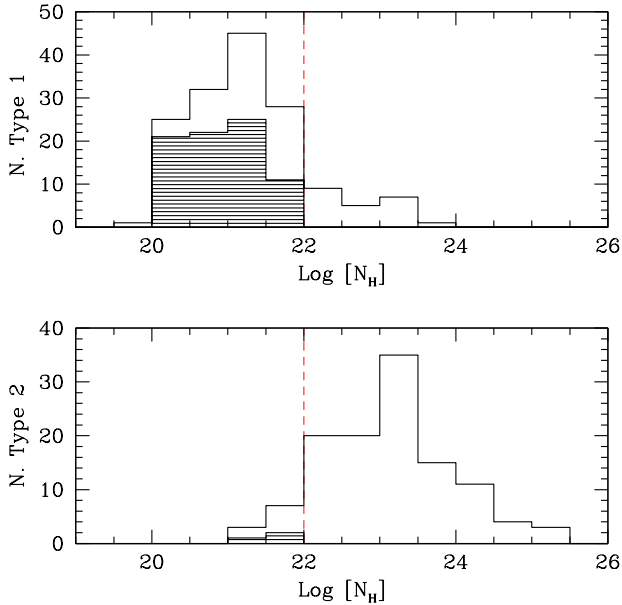}
\caption{Column density distributions in type 1 (up) and type 2 (down). Dashed bins like in figure 2.}
\label{fig1}
\end{figure}

\section{Optical classification versus X-ray absorption}
The optical classification scheme which follows the orientation-based AGN unified model (Antonucci 1993) is expected to be strictly correlated with the X-ray absorption. 
However, recently quite a few studies have claimed the existence of objects for which optical and X-ray classifications do not match. 
Although not yet explained physically, it is commonly found in X-ray selected samples that $\sim$10\% - 30\% of AGN which have only narrow lines in their optical spectra, 
suggesting extinction and thus classified as type 2, do not show absorption in their X-ray spectra (e.g. Panessa \& Bassani 2002, Tozzi et al. 2006). 
On the other hand, there are more and more objects optically classified as type 1 which show significant amount of absorption in their X-ray spectra, a 
feature which is at odds with the presence of broad lines in the optical band (e.g. Garcet et al. 2007, Panessa et al. 2008).\\
In this section we explore the connection between optical classification and X-ray absorption in the INTEGRAL AGN sample.
Before proceeding with this comparison we need however to make some assumptions.\\
Intermediate subclasses of Seyferts 1.2 and 1.5 are assigned to type 1 AGN, while those belonging to other types (1.8, 1.9) are considered as type 2 objects. 
Implicit in this, is the assumption that the late intermediate types (1.8, 1.9) are AGN viewed at intermediate  inclination angles to the central source, 
possibly through the "atmosphere" of the dusty torus. In  type 2 objects we have also included Cen B, which is a narrow line radio galaxy and
also the 5 XBONGs discussed previously; 
the underlying assumption is that in these AGN both broad and narrow line regions are hidden by gas and dust. 
Also the 2 "pure"  LINERs (MCG+04-26-006 and IGR J19118-1707) have been considered as type 2 AGN, since in their optical spectra  only narrow line components are present.  
22 objects in the sample are classified as  blazars: 8 are BL Lac objects  and  14  QSO. The former tend to be closer than the latter. 
These blazars have all been included in  type 1 AGN for the following reasons: 
a) all those classified as QSO are broad line objects;  b)  all BL Lac can be  interpreted as AGN where  a 
relativistic jet,  which  is closely aligned to the line of sight to the observer, swamps the broad and narrow line region making the source optical  spectrum featureless.\\
The overall subdivision into optical classes is summarized  in Table 1 where we also quote the percentage of type 1 (57\%) and of type 2  (43$\%$) objects.  
These percentages are not very different from those obtained following the X-ray classification: as seen previously 52\% of our objects are unabsorbed and 48\% are 
absorbed, suggesting that the AGN unification model generally holds.
However,  the match is not perfect  as we have a number of absorbed 
type 1 sources  as well as a number of unabsorbed type 2 objects.
This is clearly visible in the histograms of the column density  reported in figure 5 for type 1 (up)  and  for type 2 objects (down) separately; the line corresponding 
to N$_{H}$=10$^{22}$ cm$^{-2}$ is also plotted in order to immediately see how many sources fall in the forbidden regions (above and below this value for type 1 and 2 objects respectively). 
This is quantified in Table 2, which provides the AGN subdivision according to both  optical and X-ray classifications. 
There are 240 objects for which the match is as expected; these sources constitute  88\% of the sample. 
The remaining objects, that is 22  type 1 AGN which are absorbed  and 10  type 2 AGN which are unabsorbed, make the remaining 12\%.
In order to verify this correlation and to be able to compare our results with previous studies, the fourfold point correlation coefficient $r$ defined as in Garcet et al. (2007) has been calculated.
The use of this  coefficient represents a rigorous way to quantify the correlation between the X-ray and optical classifications: when $r$=0  no correlation is observed, while when $r$=1 there is a strict correlation. 
We found a fourfold coefficient $r$=0.77$\pm$0.05  which indicates a good agreement between optical  and X-ray classification. 
Such a high correlation value was only found by Caccianiga et al. (2004) for their sample of bright X-ray sources; 
when samples of weaker sources are used, $r$ becomes much smaller (around 0.3) implying only a mild correlation between optical and X-ray classifications 
(Garcet et al. 2007). However, as fully discussed by Garcet et al. (2007), this may be due to difficulties in dealing  with objects which are weaker or further away.

\begin{table}
\begin{center}
\caption{Number of sources as a function of the optical (type 1 and type 2) and X-ray (unabsorbed and absorbed) classifications.}
\begin{tabular}{lccc}
\hline
\hline
                         & Opt. type 1       &  Opt. type 2    &           Total \\ 
\hline
Unabsorbed     &    132                &   {\bf 10}         &  140 (52\%)  \\                
Absorbed         &   {\bf 22}           &    108              &   132 (48\%) \\
\hline
Tot                   &  154 {\bf (57\%)}              &  118 {\bf (43\%)}             &  272   \\
\hline
\hline
\end{tabular}
\end{center}
\end{table}

\begin{table*}
\begin{center}
\caption{Main characteristics of absorbed type 1 Seyferts in INTEGRAL AGN sample}
\begin{tabular}{lcccccl}
\hline
\hline
 Name                      &  class          & N$_{HOpt}$    &            N$_{HX}$     &  type of abs.$^{\dagger}$       & instr.           & ref for type of abs.  \\ 
                                &                     & $\times$ 10$^{22}$ cm$^{-2}$    & $\times$ 10$^{22}$ cm$^{-2}$ &                   &             & \\
\hline 
UGC 3142               &    Sey 1       &     -        &     4.0$^{+2}_{-1}$                       &  CA               & XMM  &   Ricci et al.  2010  \\                
IGR J21247+5059   &    Sey 1      &      -        &     7.9$^{+2.02}_{-1.66}$            &  CA                  &  XMM      & Molina et al. 2009\\    
\hline
4U 0557-385             & Sey 1.2  &    0.36$^{(1)}$    & 1.3$^{+0.2}_{-0.2}$     &  WA             & XMM & Ashton et al. 2006 \\
IGR J10147-6354     & Sey 1.2         &       -                   & 2.0$^{+1.6}_{-1.1}$     &  {\it no info}   & XRT       & Rodriguez et al. 2008 \\
MCG-6-30-15           & Sey 1.2          & 0.4-1.1$^{(2)}$  & 1.5$^{+0.3}_{-0.4}$     &    WA             & XMM   & Molina et al. 2009 \\
IGR J16558-5203     & Sey 1.2         &     -                      & 30$^{+11}_{-8}$         &     CA             & XMM      & Panessa et al. 2008 \\
IGR J19491-1035     & Sey 1.2         &      -                     & 1.8$^{+0.8}_{-0.7}$ &  {\it no info}  & XRT        & Malizia et al. 2011 \\
QSO B2251-178       & Sey 1.2         &    0.28$^{(3)}$     & 2.1$^{+0.6}_{-1.2}$               &  WA                              & Suzaku     & Winter et al. 2012\\
\hline                         
Mrk 6                       & Sey 1.5       &  0.44$^{(4)}$         & 8.12$^{+4.76}_{-2.83}$    & WA                  &   XMM       & Winter et al. 2012 \\
IGR J09253+6929   & Sey 1.5      & -                             & 14.8$^{+28}_{-11}$          & {\it no info}       & XRT          & Malizia et al. 2011 \\
NGC 3227               &  Sey 1.5     &   0.33-0.6$^{(5)}$   & 6.7$^{+0.4}_{-0.2}$          & WA                  & Suzaku      & Markowitz et al. 2009 \\
NGC 3516               & Sey 1.5      &  $<$0.07$^{(6)}$    & 3.2$^{+0.4}_{-0.4}$          & WA                   & XMM         & Mehdipour et al. 2010 \\
NGC 3783               & Sey 1.5      &  0.18$^{(7)}$          & 1.2$^{+0.4}_{-0.4}$          & WA                   & Suzaku      & Winter et al. 2012 \\
NGC 4151               & Sey 1.5      & 0.13-0.43$^{(8)}$   & 21.4$^{+13.7}_{-10.4}$   & WA                    & Suzaku      & Winter et al. 2012 \\
IGR J12107+3822   & Sey 1.5      & -                             & 4.6$^{+1.2}_{-1.0}$          & {\it no info}        & XMM          & Parisi et al. 2012 \\
4U 1344-60             & Sey 1.5      & -                              & 46.8$^{+32.4}_{-21.2}$    & CA                    & XMM          & Molina et al. 2009 \\
ESO 140-43            & Sey 1.5       & 0.19$^{(9)}$          &   11$^{+3}_{-6}$               & WA                    & XMM         & Ricci et al. 2010 \\
Swift J1930.5+3414 & Sey 1.5     &  -                             & 27.5$^{+20.4}_{-13.7}$    & CA                    & XRT          & Winter et al. 2009 \\
\hline
IGR J16185-5928    & NLS1         &   -                           & 10.6$^{+16}_{-5.5}$          &   CA                & XMM         & Panessa et al. 2011\\
IGR J19378-0617    & NLS1        & 0.22-0.64$^{(10)}$  & 32$^{+10.5}_{-8}$           & CA                    & XMM        & Panessa et al. 2011 \\           
\hline
\hline
\end{tabular}
\end{center}
${\dagger}$: CA = complex absorption i.e. multi absorption layers of neutral material; WA = warm absorber i.e. multiple ionized absorbers with different covering factors, 
column densities and ionization parameters (in both cases we have considered the maximum value of N$_{H}$); no info = no information on the type of absorption is available. 
(1) Turner et al. 1996; (2) Pounds et al. 1986; (3) Wu et al. 1980; (4) Feldmeier et al. 1999; (5) Kraemer et al. 2000; (6) Kraemer et al. 2002 ; (7) Word \& Morris 1984; 
(8) Shapovalova et al. 2010; (9) De Zotti  \& Gaskell 1985 ; (10) Mullaney \& Ward 2008
\end{table*}

\subsection{Absorbed type 1 AGN}
Among type 1 AGN, 22 objects are absorbed including  2 Seyfert 1s, 6 Seyfert 1.2, 10 Seyfert 1.5,
2 NLS1 and 2 blazars, corresponding to $\sim$14\% of the  entire type 1 AGN population. Using the errors on column density reported 
in  the table in the Appendix, we can define with some precision that the fraction of absorbed type 1 AGN ranges from 12\% to 17\%.
This percentage is greater than the $\sim$11\% found by Garcet et al. (2007) in their XMM selected sample.
These authors also found  that their absorbed type 1 AGN were  high X-ray luminosity objects lying at high redshift (typically above z=1)
giving an indication that this may be a common property at high redshift and luminosity values.
This result, which has been recently confirmed by Mateos et al. (2010) and Scott et al. (2011) using 2-10 keV data,
is not found in our hard X-ray sources: in fact, although our objects are relatively  luminous in X-rays, they are not brighter than other type 1 
objects nor lie at higher redshift (see table in the Appendix).\\
Several interpretations have been suggested in order to explain the nature of absorbed type 1 AGN.
For example, Maiolino et al. (2001a) have discussed a sample of nearby AGN whose X-ray spectra show evidence for cold absorption but there was no  hint of obscuration 
in their optical data (hence their classification as type 1). They concluded that the  ratio A$_{V}$/N$_{H}$ in these objects is systematically much lower than the Galactic standard value. 
In a companion paper, Maiolino et al. (2001b) suggested that a dust distribution dominated by large grains in the obscuring torus could explain the low 
A$_{V}$/N$_{H}$ values obtained. They claim that the formation of large grains is naturally expected in the high density environment characterizing the circumnuclear region of AGN. 
These large grains make the extinction curve flatter than the Galactic one and thus for a given N$_{H}$ value  a reduced extinction and reddening are observed, 
compared to the Galactic standard. \\
An alternative explanation was  proposed by Weingartner \& Murray (2002): they  suggested that the line of sight to these AGN passes through
ionized material located just off the torus and/or accretion disk.
This material is responsible for the X-ray absorption, while the optical/infrared extinction occurs in material farther from the nucleus,
where the dust may be quite similar to the Galactic dust. 
The X-ray-absorbing material may be dust-free or may contain large grains that have very small extinction efficiencies in the optical/infrared. 
This material may be associated with a disk wind, which would originate within the dust sublimation radius (see Murray et al. 1995). 
In this case, the dust will sublimate and the obscuration/extinction in the optical will be much reduced, even
if there is a strong absorption in the X-rays, produced by the ionized gas.\\
Finally variability, i.e. non simultaneous X-ray and optical observations, could also explain the apparent discrepancy between the optical and X-ray classifications of some type 1
AGN.\\
In order to study these absorbed type 1 AGN, we analysed in detail the available X-ray data  of each object to have
more information on the nature of the absorption measured.
The 2 absorbed blazars will not be considered in the following discussion since, as showed in section 4.2, in these objects (PKS 1830-211, De Rosa et al. 2005 and 
IGR J22367-1231, Bassani et al. 2007) the absorption is probably not intrinsic. \\
We have reported in Table 3  the remaining {\bf 20} absorbed type 1 AGN, together with their specific optical sub-class,  intrinsic dust reddening  when available, X-ray column density, 
type of  X-ray absorption (complex  or ionized) reported in the literature, satellite used for the  X-ray measurement and {reference for the type of absorption}.
The dust reddening expressed as N$_{H_{Opt}}$ has been estimated from the  E$_{B-V}$ intrinsic to the source assuming the Cardelli et al. (1989) extinction law and 
N$_{H_{Opt}}$= 2.22 $\times$ 10$^{21}$ A$_{V}$, where A$_{V}$=3 $\times$ E$_{B-V}$ (Zombeck 1990).
It is worth noting that in those cases where the absorption was found to be variable or complex  the highest N$_{H}$ value has been preferred. \\
As evident from a comparison between the optical and X-ray column densities, dust extinction tends to be  systematically lower than gas absorption, in agreement with 
Maiolino et al. (2001a), implying that a non standard  dust to gas ratio is a viable explanation for the absorption in our type 1 objects. 
A more interesting aspect emerging from Table 3 is the fact that many objects have X-ray spectra characterized by either complex (CA) i.e. multi absorption layers of cold material or warm/ionized (WA) absorption. 
These two models are somehow interchangeable as demonstrated for MKN 6 by Schurch et al. (2006) and Page et al. (2011), implying that when  absorption is found in type 1 AGN it  
can be modeled in both ways. The above authors also suggested that, between CA and WA, the second is to be preferred since  it offers a more physical and  testable 
description of the X-ray data; the WA model also  provides a more natural way to explain the variability in the absorption seen in many objects as due to the presence 
of several distinct ionization phases. A possible location for this ionizing gas could be in an accretion disk wind or in the biconical structure mapped by the [O III] 
emission line (ionization cones) and seen in some objects, including NGC 4151, MKN 6, NGC 3227 and  NGC 3516.  
The Chandra survey of Extended Emission-line Regions in nearby Seyfert galaxies (CHEERS project) provides 
some support to the  ionization cones hypothesis (Wang et al. 2011, Paggi et al. 2012).\\
As can be seen in Table 3,  for a few cases we do not have detailed information on the nature of absorption, but only a
simple neutral column density in excess to the Galactic value.    
However, for IGR J10147-6354, IGR J19491-1035 and IGR J09253+6929, the N$_{H}$ estimates are based on  rather short (not more than 6 ksec) Swift-XRT observations
and so it would be important to have higher quality spectral information  before assessing the type of absorption in these sources. 
We have also re-analysed the Newton-XMM data of IGR J12107+3822 taken from the work of Parisi et al. (2012) which lists a number of objects fitted with a standard basic model. 
Unfortunately the data are not of high quality, the source remains absorbed in the various models tested, 
and  we cannot exclude at the present stage whether the absorption is simple or rather complex/ionized.\\
Regarding the possibility that optical/X-ray mismatch is due to variability, we note that a few of the sources listed in Table 3 have changed their classification in time: NGC 4151 from type 1 to type 2 and back,
MKN 6 and NGC 3227 from type 2 to type 1 (Osterbrock \& Koski 1976, Shapovalova et al. 2009 and Pronik 2009). However, these type of transitions have always  been
associated to continuum changes with the type 2 designation found when the optical ionizing flux was in a low state. Thus in such objects, the weakness of the broad lines
can be entirely independent of obscuration/reddening i.e. these remain type 1 AGN even if temporarily classified as type 2.\\
In conclusion, we find that  no more than 20 objects, corresponding to  13$^{+3}_{-2}$\% of our type 1 sample, have 
N$_{H}$ $>$ 10$^{22}$ cm$^{-2}$. This percentage has to be considered as an upper limit; the lower limit can be put at 11\% by excluding  the 3 sources for which only 
short XRT exposures are available.
This percentage range coincides with that found by Garcet et al. (2007).\\
We also find that some  type 1 AGN in Table 3 have dust-to-gas ratio different than the Galactic one and all have a similar type of absorption. 
Therefore both explanations put forward to account for absorbed type 1 AGN  can work, although the warm ionization hypothesis seems at this stage favoured. 
In any case, the absorption in these objects seem to be unrelated to the  toroidal structure invoked by the AGN unifying theory and 
so absorbed type 1 sources do not seem to question its validity. 

\begin{table*}
\begin{center}
\caption{Main characteristics of unabsorbed type 2 Seyferts in INTEGRAL AGN sample}
\begin{tabular}{lcccccl}
\hline
\hline
 Name                 &  class      & N$_{HOpt}$      &             N$_{HX}$                       &  type of abs.       & instr.           &     ref \\ 
                             &               &   $\times$ 10$^{22}$  cm$^{-2}$                & $\times$ 10$^{22}$  cm$^{-2}$ &                              &             & \\
                             \hline
NGC 5995                & Sy 1.9       & 1.18$^{(1)}$    &   0.85$^{+0.41}_{-0.29}$                                  & intr                      & ASCA       & Shu et al 2007 \\
IGR 17513-2011      & Sy 1.9      &   -                       &   0.66$^{+0.01}_{-0.01}$    & intr                      & XMM         & De Rosa et al. 2012\\
IGR J19077-3925    & Sy 1.9     &  0.07$^{(2)}$ &    0.14$^{+0.10}_{-0.08}$        & intr                     & XRT           & Malizia et al. 2011 \\
\hline 
IGR J01545+6437 & Sy 2       &  0.25$^{(2)}$   &     0.66                                    & Gal                     & XRT           & Malizia et al. 2011 \\
IGR J02504+5443 & Sy 2       &  0.30$^{(3)}$    &     0.42                                   & Gal                     & XRT         & Landi et al. 2007 \\
IGR J03249+4041(SW)$^{\dagger}$  & Sy 2      &   1.32$^{(4)}$  &     $>$0.15    & intr                 & XRT          & Malizia et al. 2011\\
IGR J07565-4139   & Sy 2      & $>$0.9$^{(5)}$ &    0.72$^{+0.04}_{-0.04}$   & intr                      & XMM         & De Rosa et al. 2012 \\
NGC 2992                & Sy 2     & 0.47$^{(6)}$    &      0.80$^{+0.05}_{-0.01}$                                 & intr                      & Suzaku     & Yaqoob et al. 2007 \\
IGR J14515-5542   & Sy 2       &  -                   &    0.33$^{+0.17}_{-0.08}$        & intr                      & XMM         & De Rosa et al. 2012 \\
IGR J16024-6107   & Sy 2       &   0.10$^{(3)}$    &    0.25 $^{+0.03}_{-0.01}$   & intr                      & XMM         & De Rosa et al. 2012 \\
\hline
\hline
\end{tabular}
\end{center}
$\dagger$  interacting galaxies (see text for details and table in the appendix).\\
(1) Lumsden et al. 2001; (2) Masetti et al. 2010; (3) Masetti et al. 2008b; (4) Masetti et al. 2012; (5) Masetti et al. 2006; (6) Trippe et al. 2010
\end{table*}

 \subsection{Unabsorbed type 2 AGN}
Ten  objects in the sample (see Table 4) are not absorbed in X-rays despite being classified as type 2 AGN. 
However, 3 of these are of intermediate type and so display evidence for the presence of a broad line region and therefore,
following the assumptions we made in section 5, they are expected to be absorbed in X-rays more than type 1, although not as  much as type 2 AGN.  
Indeed the  column density distribution for type 1.9 and that for type 2  indicate slightly less absorption in intermediate type objects (see also Risaliti et al. 1999).
At least 2 (NGC5995 and IGR J17513-2011) of the 3   Seyfert 1.9 which are unabsorbed in X-rays 
have a column density close to 10$^{22}$ cm$^{-2}$. Taking into consideration  the large uncertainties often associated with the measurement of N$_{H}$, these objects can still fit 
with the unified theory if our line of sight grazes the outer edge of a central obscuring torus.
However, this is not the only possible interpretation as intermediate classifications may also be related to other phenomena such as an intrinsically variable 
ionizing continuum or the presence of absorption/reddening unrelated to the torus: for example 
a source that would normally appear as a Seyfert 1 can be classified as an intermediate type if it is in a low  optical flux state (Trippe et al. 2010) or
a source could have the BLR obscured (except for the strongest H$\alpha$ line) by dust related to  large scale structures such as bars, dust lanes, and host galaxy plane
(Malkan et al. 1998). Unfortunately we do not have sufficient information on these 3 sources to test either hypothesis although we note that
NGC 5995 can be classified as an HII galaxy based on some diagnostics and as a Seyfert based on others (Shi et al. 2010), while 
IGR J19077-3925 is possibly interacting with a nearby object: it is possible that dust unrelated to the 
torus is present in both AGN in the form of a starburst or due to interaction.\\
The remaining 7 objects are Seyfert 2 not absorbed in X-rays: they make a small fraction (from 6 to  8\% allowing for errors on N$_{H}$) of all type 2 AGN  in the sample. 
Panessa and Bassani (2002) estimated the percentage of this type of source to be in the range 10$\%$-20$\%$; this number which is derived from a non complete sample,
is higher than that found by Risaliti et al. (1999) in a sample of optically selected Seyfert 2 (4$\%$)
but consistent with the estimate (12$\%$) made by Caccianiga et al. (2004). Much larger fractions (66-68$\%$) are reported by Page et al. (2006) and Garcet et al. (2007).  
Despite the above uncertainties,  unobscured type 2 AGN may be  a non-negligible part of the AGN population and one which mostly questions  
the AGN unification theory based purely on orientation: they may harbour a genuinely weak or absent broad line region (hence the name of  "true" type 2 AGN), 
and thus are not the simple obscured version of type 1 objects expected from the unified model. \\
However, before claiming that our 7 sources are true type 2 AGN we should exclude alternative interpretations.
For example it has been argued that some "unobscured" Seyfert 2 galaxies are actually Compton-thick objects in which the direct nuclear component below 10 keV 
is completely suppressed and we would only witness an unabsorbed spectrum due to scattered nuclear radiation and/or host galaxy emission from a circumnuclear starburst; 
the presence of heavy obscuration can also be  lost in a low S/N X-ray spectrum.
A clear signature of the Compton thick nature of a source is the presence of a strong FeK$\alpha$ line with an equivalent width of $\sim$1 keV: none of the sources with good 
quality X-ray data (ASCA, XMM and Suzaku) display such an evidence. On the other hand, XRT spectra do not have sufficient sensitivity to allow the detection of the iron line 
in weak objects.\\
Another method to recognize such misclassified Compton thick sources, is to use the diagnostic diagram of Malizia et al. (2007), 
which plots the X-ray absorption as a function of the source  flux  ratio F$_{(2-10)keV}$/F$_{(20-100)keV}$: for our sample of type 2 AGNs this is shown 
in  figure 6 where Log N$_{H}$ is the value reported in the table in the Appendix and which is  either the intrinsic or the Galactic absorption,
this latter being  taken again as an upper limit to the  X-ray column density. 
A clear trend of decreasing flux ratios as the absorption increases is expected due to the fact that the 2-10 keV flux is progressively depressed as the
absorption becomes stronger. Indeed  the two lines shown in the figure  describe how the flux ratio changes as a function of N$_{H}$ 
in the case of objects characterized by  an absorbed power law having a photon index of 1.5 and 1.9 respectively. 
It is evident that the majority of our sources  follow the expected trend with most absorbed  AGN showing progressively 
lower F$_{(2-10)keV}$/F$_{(20-100)keV}$ values.  
Misclassified Compton thick AGN are expected to lie in the region of low X-ray flux ratio and low N$_{H}$ values. 
Four  type 2 AGN: LEDA 96373 (\#3), IGR J12288+0052 (\#4 a source already discussed in section 4.2), IGR J18311-337 (\#5) 
and IGR J01545+6437 (\#6) are located in this region but, while three are  absorbed, although not heavily,
the only one with N$_{H}$  below 10$^{22}$ cm$^{-2}$  is IGR J01545+6437.\\
For this last source  the F$_{2-10~keV}$/F[O{\sc iii}] ratio\footnote{The [O{\sc iii}] flux is corrected for extinction using the prescription of Bassani et al. (1999a)} often used as another  way to pinpoint 
heavily absorbed Seyfert 2,  is small (0.64) but not so much as to unambiguously  classify the source as a Compton thick object (Bassani et al. 1999a). 
Clearly in this case, where only poor quality X-ray data are available, a statistically more significant X-ray spectrum is needed 
to better classify this source in terms of absorption. The lack of deep X-ray observations is also the reason that prevents us to give conclusive results in the cases of 
the other 3 sources (for example the detection of strong iron lines) which, although absorbed, have a low X-ray softness flux ratio; indeed  also the F$_{2-10}$ keV/F[O{\sc iii}] ratio is  not sufficiently constrained to discriminate between Compton thin and Compton thick absorption.\\
Various  other explanations have been proposed to understand the nature of unabsorbed  type 2 AGN such as
state transitions and non-simultaneous X-ray and optical observations, as already mentioned. Some AGN change their optical classification over time, while  
others display rapid variations in their X-ray column density: such objects might be classified as unobscured (in X-rays) type 2 objects by chance, 
if the X-ray and optical data are obtained at different times. One example of such an object is NGC 2992. This source has a long history of  variability
both in the optical and X-ray regime (Trippe et al. 2008): it was originally classified as a Seyfert 1.9, but it has been seen previously as a Seyfert 1.5 with strong broad H$\alpha$
emission while more recently it changed again to a type 2 AGN.  Because the changes in the optical seem to be correlated with the X-ray brightness, it seems that NGC 2992 is a type 1
Seyfert (hence unabsorbed) becoming a type 2  AGN only when the nucleus is in a low continuum state and the broad emission lines are extremely weak or absent. 
The small absorption present seems to be related to a dust lane in the source rather than to an obscuring torus (Trippe et al. 2008). 
It is difficult to assess if a similar behaviour is present in other unabsorbed Seyfert 2 galaxies in the sample since we lack a long history of observations for those AGN which have only 
recently been discovered by INTEGRAL. 
Nevertheless, for a few objects in Table  4, we find  evidence for variability at least in X-rays that could indicate a similarity with NGC 2992.
IGR J16024-6107 is seen varying in  the INTEGRAL data being a burst source (see Bird et al. 2010 for details)  and also between Swift XRT
observations (the count rate in the 0.3-10 keV band dropped by a factor of 1.7 over 21 days period).\\
Also IGR J07565-4139 is seen varying between XRT, Chandra and XMM observations (Malizia et al. 2007, De Rosa et al. 2008, De Rosa et al. 2012) while  IGR J14515-5542 shows a different flux  when  two XMM Slew observations performed 7 months apart are compared. \\
It is thus important for these objects to perform optical and X-ray spectroscopy as close in time as possible.\\

\begin{figure}
\includegraphics[width=1.0\linewidth]{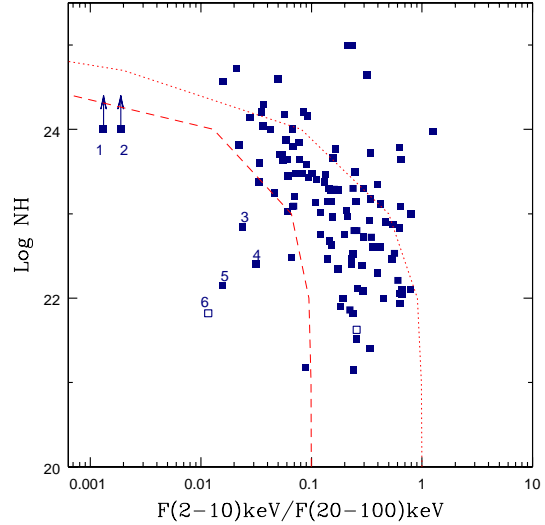}
\caption{Log N$_{H}$ versus F$_{2-10~keV}$/F$_{20-100~keV}$ flux ratio of the INTEGRAL type 2 AGN. 
Open symbols are objects where no intrinsic absorption have been measured and lines correspond to expected values for an absorbed power law with 
photon index 1.5 (dot) and 1.9 (dash).
Sources: \#1=IGR J14561-3738, \#2=MCG-07-06-018,  \#3=LEDA 96373, 
\#4=IGR J12288+0052, \#5=IGR J18311-337, \#6=IGR J01545+6437}
\label{fig1}
\end{figure}

Another explanation put forward to interpret  unabsorbed type 2 AGN is an
extremely high dust-to-gas ratio (N$_{HOpt}$/N$_{H}$) compared with the Galactic value.
This option has been poorly explored in the literature because the few measurements so far available
indicate that this ratio is generally 3-100 times lower than the Galactic  value (Maiolino et al. 2001a).
This is basically the reverse situation described in the previous section for absorbed type 1 AGN where  observations
require more gas than dust; in this case instead  we need less absorbing gas and much more optical dust reddening.
Also in this case, we have been able to collect E$_{B-V}$ values intrinsic to the AGN for most of the sources in our  sample  (see column 3 of Table 4) and have converted them into an optical column density as done in Table 3.
In most cases, the optical (dust) absorption  is similar within errors to the  X-ray (gas) one, implying a dust to gas ratio in the range 0.4-1.4, i.e. similar to the Galactic value, an
indication that in these cases this  explanation is not appropriate. 
The only possible exception is IGR J03249+4041(SW), which has an optical reddening of
1.3 $\times$10$^{22}$ (Masetti et al. 2012) a factor of 10 higher than the admittedly loose  lower limit on the X-ray column density. 
This suggests that either the X-ray absorption is largely underestimated or that dust is present in large quantities in the source and is responsible for  masking 
the broad line region. IGR J03249+4041(SW) is part of a group of 3 galaxies in a common halo (Lutovinov et al. 2010);  at least two galaxies are AGN (IGR J03249+4041(SW) and IGR J03249+4041(NE);
see also the table in the Appendix) interacting among themselves, which gives rise to a complex structure (Meusinger et al. 2000).
It is therefore not surprising to find such large quantities of dust in this galaxy.\\
The last possible explanation for unabsorbed type 2 AGN is the dilution effect: broad emission lines in these objects can be overwhelmed by the host galaxy light in addition to being hidden in a low  S/N optical spectrum. 
However, this effect is unlikely to be important in the spectra acquired within our own optical identification program since their S/N ratio is good enough to characterize the shape of the 
emission lines. Also the spectral  resolution is sensitive enough to definitely assess wether a line is narrow or broad; therefore we do nor expect that the contribution of the underlying 
continuum could alter the classification (Masetti et al. 2012 and references therein).\\
Finally, it is worth noting that in a  few objects of Table 4 the X-ray data are of low statistical significance and of limited spectral coverage
(mainly from XRT observations); in these cases it is possible that more refined measurements will
provide a more precise estimate of the source  column density and a value more in line with the object optical class.\\
If a source cannot be explained by any of the above hypotheses, then it is probably a  "true" type 2 AGN, i.e.  a source with a very weak or even  absent BLR.
After close scrutiny, only a handful of objects have been identified with true unobscured type 2 AGN in the literature (Shi et al. 2010); 
the main contamination comes from Compton thick AGN not recognized as such for lack of good quality X-ray data or broad band X-ray spectral coverage (Shi et al. 2010) 
or by type 1 AGN with optical spectra overwhelmed by a luminous host galaxy (Garcet et al. 2007).  
Our set of sources have not been studied in detail to exclude these alternative explanations.
Furthermore, although our unabsorbed type  2  AGN tend to be dimmer  in X-rays   than absorbed ones, their  2 --10 (and 20 --100) keV luminosities are  always above 10$^{42}$ erg s$^{-1}$:
this is incompatible with models (Nicastro 2000; Nicastro et al. 2003; Elitzur \& Shlosman 2006)  where  at low accretion rates (and hence
low luminosities)  the BLR cannot form.\\
Therefore we can estimate that  at most only half of the sample is  still eligible 
to be classified as a "true" type 2 AGN, i.e. 3-4$\%$ of the entire type 2 poputation, a value similar to the original estimate by Risaliti et al. (1999).

\section{Summary and Conclusions}
In this work we have presented a new INTEGRAL AGN catalogue which, being made by collecting all the AGN from the 4th IBIS catalogue and the Krivonos et al. (2007)
IBIS sky survey, represents the most complete view of the INTEGRAL extragalactic sky up to now. 
It lists 272 AGN for which we have secure optical identifications, precise optical spectroscopy and  measured redshift values plus X-ray
spectral information, i.e. 2-10 keV and 20-100 keV  fluxes and  column density. \\
A list of AGN properly characterized in this way is an ideal tool for population studies. \\
Here  we mainly used this sample to study the absorption properties of active galaxies, to probe new AGN classes and to test the AGN unification scheme; 
some of the results presented here are the refinement of previous works, while others are
obtained here for the first time using a hard X-ray selected sample.\\
Assuming as a dividing line between absorbed and unabsorbed AGN a column density of 10$^{22}$  cm$^{-2}$, we find that half (48\%) of the sample is absorbed while the fraction of 
Compton thick AGN is small ($\sim$7\%). In line with our previous analysis,  we have  shown that these fractions suffer from a bias towards heavily absorbed objects 
which are lost if weak and at large distance. When this bias is  removed, as is possible in the local Universe, the above fractions 
increase to  80\% and 17\%  (within 60 Mpc) in full agreement with our previous results (Malizia et al. 2009). 
We also find that absorption is a function of source luminosity, which implies some
evolution in the obscuration properties of AGN. Unfortunately it is difficult at the present stage to disentangle between this  evolution effect and the bias discussed above
as the two are strongly correlated. This important  aspect will be explored in the future with a dedicated study. \\
Among the 272 AGN,  a few peculiar classes, so far poorly studied in the hard X-ray band, have been detected for the first time such as 5 XBONG, 5 type 2 QSOs and 11 LINERs.
The properties of the 5 XBONG  can be explained in terms of heavy obscuration  combined with dilution due to a bright galaxy host; 
given the X-ray luminosity involved, these objects are unlikely to be explained in terms of radiatively inefficient accretion. 
None of the type 2 QSO is heavily absorbed nor qualifies to be a ERO;
type 2 QSO make a small fraction of the sample (2\%) and are  largely outnumbered by type 1 QSO, suggesting that they are less numerous or more difficult to find. 
Finally we have been able to observe hard X-ray selected LINERs for the first time: akin to the Seyferts they come in two flavours (type 2 and 1) and, similarly, the first are 
absorbed while the second are not. All our LINERs are obviously powered by an AGN.\\
The careful classification of  each AGN listed in the catalogue has  allowed  the study of the correlation between optical 
classification and X-ray absorption  and hence to test the AGN  unification model.  Although the presence of a correlation is expected and indeed found, 
i.e. type 1 AGN are typically unabsorbed while type 2 AGN are often  absorbed, its strength changes from sample to sample but  is 
never 100$\%$. The outliers are clearly  very interesting objects because they question the validity of the Unified Theory and therefore can provide information on how to refine it.\\
In terms of  optical classification, our sample contains  154 objects (57\%)  which are of  type 1 and  118 (43\%) which are of type 2; this subdivision is 
similar to that found in X-rays if  unabsorbed versus absorbed objects are considered, suggesting that the match between optical and X-ray classification is good overall.
Only a small percentage of sources  (12\%) does not fulfill the expectation of the Unified Theory as we find 22 type 1 AGN which are absorbed and 10 type 2 AGN 
which are unabsorbed .\\
Looking in depth at these sources we conclude that:
\begin{itemize}

\item Most of the absorbed type 1 AGN have dust extinction  systematically lower than gas absorption, confirming previous results from Maiolino et al. (2001a).
More interestingly, however, is the fact that many absorbed type 1 AGN have X-ray spectra characterized by either complex  or warm/ionized absorption; these two models
provide quite similar fits so that we can conclude that all absorbed type 1 AGN present the same type of absorption, more likely due to ionized gas located in an
accretion disk wind or in the biconical structure associated to the central nucleus. Since this type of absorption is unrelated to the toroidal structure invoked by the AGN unifying theory, 
absorbed type 1 sources do not seem to question its validity.\\

\item We also analysed all  10 type 2 AGN which resulted to be  unabsorbed to see if  the lack of  X-ray absorption could be explained somehow before considering them  as "true" 
type 2 AGN. We found that a) one is possibly a Compton thick object (IGR J01545+6437), b) a few sources  are variable in X-rays so that their different optical/X-ray 
classifications can be explained in terms of state transitions and/or non-simultaneous X-ray and optical observations (NGC 2992, NGC 5995, IGR J16024-6107), c) at least one source could be
characterized by an extremely high dust-to-gas ratio (IGR J03249+4041). All together we estimate that at most only half of the sample is still eligible to be classified as a "true" 
type 2 AGN, i.e. 3-4\% of the type 2 sub-sample.
\end{itemize}

In conclusion, the standard-based AGN unification scheme is followed by the majority (up to 96--97\%)of bright AGN, very few outliers are found among type 2 AGN and almost none among type 1 sources; 
despite absorption being present in a significant fraction of type 1, it is likely unrelated to the torus but rather to ionized gas near the source central engine.\\
These are some of the results that can be obtained with the present large sample, highlighting the
potential of statistical and population studies using hard X-ray selected AGN. \\
Further work involving absorption on a large scale is in progress.

\section*{Acknowledgments}
We thank the anonymous referee for her/his valuable comments and suggestions.\\
This research has made use of data obtained from the SIMBAD database operated at CDS, Strasbourg, France; 
the High Energy Astrophysics Science Archive Research Center (HEASARC), provided by NASA's Goddard Space 
Flight Center NASA/IPAC Extragalactic Database (NED). 
The authors acknowledge the ASI financial support via ASI--INAF grants I/033/10/0 and I/009/10/0.
%We thank the anonymous referee for useful 
%remarks which helped us to improve the quality of this paper.

\appendix
\section{INTEGRAL/IBIS AGN sample}
In this Appendix we report the table containing the optical coordinate, redshift, class, 20-100 keV flux  and X-ray data (2-10 keV flux, 
column density and reference work from which these data ) for the full AGN sample presented and discussed in this work.\\
\vskip0.8cm
{\bf NOTES on table A1}: Names in bold are for hard X-ray sources from the INTEGRAL all-sky survey by Krivonos et al. 2007 (see text).
\begin{list}{}
\item F$_{H}^{\dagger}$ = F$_{20-100~keV}$ in units of 10$^{-11}$ erg cm$^{-2}$ s$^{-1}$
\item F$_{S}^{\dagger\dagger}$ = F$_{2-10~keV}$ in units of 10$^{-11}$ erg cm$^{-2}$ s$^{-1}$
\item ${\diamond}$ value of N$_{H}$ in bold are Galactic column density measured in the source direction (see text)
\item $^{\star}$ flux variability
\item $^{(a)}$ interacting galaxies where the INTEGRAL/IBIS detection is referred to both galaxies and the 20-100 keV flux has been estimated to be
50\% to each galaxy (see text for details).
\item $^{(b)}$ this source has a wrong name (ESO 548-G01) in the 4th IBIS catalogue, here the right name and the right redshift has been reported
\item $^{(c)}$ a different value of column density comes from a published Chandra observation (Tomsick et al. 2007) where it is not clear wether the authors include Galactic absorption or not in their estimate of N$_{H}$.
This source has now been observed  by XRT and these more recent data provide a column density 
of N$_{H}$ = 0.95$^{+0.34}_{-0.30}$ $\times$ 10$^{22}$ cm$^{-2}$; since the Galactic absorption in the source direction is $\sim$ 10$^{22}$ cm$^{-2}$ it is very likely that the Chandra estimate includes 
both Galactic and intrinsic absorption.
\item $^{(d)}$ INTEGRAL flux comes from the whole map and has to be considered as a lower limit  since the source has been best detected in a revolution map likely during a flare. Its variability and its black hole
mass (Masetti et al. 2010) are typical of hard X-ray selected blazars, however the lack of radio detection poses still some doubts on the blazar nature.
\item $^{(e)}$  ESO 138-1 and NGC 6221 are blended in the IBIS maps and trough a simulation we have  estimated the contribution of each source in the 20-100 keV band which is almost 50\%
\item $^{(f)}$ for this source we have very recently obtained XMM data. The preliminary analysis indicates that  the column density is 0.81$^{+0.05}_{-0.04}$ $\times$ 10$^{22}$ cm$^{-2}$;
this value is compatible with that measured by XRT (Molina et al 2009) but the uncertainty is now sufficiently small to consider the source as not absorbed.
\item $^{(g)}$ pair of galaxies MCG+04-48-002 (\#1) and NGC 6921 (\#2) where the INTEGRAL/IBIS detection is referred to both galaxies and the 20-100 keV flux of each has been estimated.
\end{list}

\clearpage

\begin{table}
\rotcaption[]{INTEGRAL/IBIS  AGN}
\centering
\begin{sideways}
\begin{tabular}{l l l  c l l l c l }
  \hline\hline
 Name                                      &      RA              &            Dec          &  $z$          &   Class          & F$_{H}^{\dagger}$  & F$_{S}^{\dagger\dagger}$ & Log N$_H^{\diamond}$  & Reference\\
\hline
IGR J00040+7020                & 00 04 01.92   & $+$70 19 18.5  &   0.096     &    Sy2             &   1.47    &    0.35           &  22.52 [22.49 -- 22.54]   & De Rosa et al. 2012    \\      
 IGR J00158+5605               & 00 15 54.19   & $+$56 02 57.5   &   0.169     &    Sy1.5          &   $<$0.66   &   0.31      &  {\bf 21.50} & Malizia et al. 2011     \\
IGR J00256+6821                & 00 25 32.50   & $+$68 21 44.0   &   0.012     &    Sy2              &   1.47   &   0.05           &  23.60  [22.92 -- 23.93]  & Landi et al. 2007a     \\      
IGR J00333+6122                & 00 33 18.34   & $+$61 27 43.3   &   0.105     &    Sy1.5          &   1.38    &   0.68           &  21.93 [21.65 -- 22.11]   & Molina et al. 2009       \\      
1ES 0033+595                      & 00 35 52.60   & $+$59 50 05.0   &   0.086     &    BL Lac        &   1.83   &    5.90           &  21.55 [21.44 -- 21.64]    &  Donato et al. 2005\\
IGR J00465-4005                  & 00 46 20.68   & $-$40 05 49.1    &   0.201     &    Sy2              &   3.56    &  0.12          &  23.38 [23.25 -- 23.52]   & Landi et al. 2010a   \\      
MKN 348                               & 00 48 47.10   & $+$31 57 25.0   &   0.015     &    Sy2              &  10.60   &  0.44           &  23.02  [22.93 -- 23.10]   & Tartarus Database         \\
\hline
\end{tabular}
\end{sideways}
\end{table} 

\clearpage

\begin{table}
%\rotcaption[]{continued}
\centering
\begin{sideways}
 \begin{tabular}{l l l  c l l l c l }
  \hline\hline
 Name              &      RA     & Dec          &  $z$  &   Class   & F$_{H}^{\dagger}$  & F$_{S}^{\dagger\dagger}$ & Log N$_H^{\diamond}$  & Reference\\
\hline
4C03.8                                   & 04 07 16.45   & $+$03 42 25.9   &   0.089      &    Sy2                &   3.45   &   0.21  &  23.45 [23.40 --23.47]   &  Winter et al. 2009\\               
3C 111                                   & 04 18 21.28   & $+$38 01 35.8   &   0.0485    &    Sy1                &  10.40  &   3.51  &  21.66 [21.63 -- 21.69]    & Molina et al. 2009\\               
{\bf IGR J04221+4856}          & 04 22 00.50   & $+$48 56 04.0   &   0.114      &   Sy1               &   1.42   &   0.37  &   {\bf 21.85}                     & this work  \\
LEDA 15023                          & 04 23 40.80   & $+$04 08 03.0   &   0.045       &    Sy2               &   1.87   &   0.13  &  23.48  [23.32 -- 23.59]   & Malizia et al. 2011 \\      
3C120                                   & 04 33 11.09   & $+$05 21 15.6   &   0.033       &    Sy1.5             &   7.98   &  4.60   &  21.20  [21.19 -- 21.22]    & Grandi et al. 1997 \\       
UGC 3142                             & 04 43 46.89   & $+$28 58 19.0    &   0.0216    &    Sy1               &   4.87   &  1.70   &  22.60  [22.48 -- 22.78]    & Ricci et al. 2010 \\      
LEDA 168563                       & 04 52 04.79   & $+$49 32 44.6   &   0.029       &    Sy1                &   5.55  &   4.52   &  {\bf 21.73}                        & Molina et al. 2009\\
\hline
\end{tabular}
\end{sideways}
\end{table} 

\clearpage

\begin{table}
%\rotcaption[]{continued}
\centering
\begin{sideways}
 \begin{tabular}{l l l  c l l l c l }
  \hline\hline
 Name              &      RA     & Dec          &  $z$  &   Class   & F$_{H}^{\dagger}$  & F$_{S}^{\dagger\dagger}$ & Log N$_H^{\diamond}$  & Reference\\
\hline
IGR J09026-4812$^{(c)}$  & 09 02 37.31   & $-$48 13 33.9     &  0.0391     &    Sy1             &   2.42  &   0.46    &  21.98 [21.81 -- 22.11]         &  this work \\         
SWIFT J0917.2-6221       & 09 16 09.41   & $-$62 19 29.5    &  0.0573     &    Sy1             &   1.91   &   1.43    &  21.61 [21.43 -- 21.73]        &   Molina et al. 2009 \\      
MCG-01-24-012               & 09 20 46.31   & $-$08 03 21.9    &  0.0196     &    Sy2             &   4.06   &   1.00    &  22.80 [22.76 -- 22.83]        &  Shinozaki et al. 2006 \\
Mrk 110                              & 09 25 12.85   & $+$52 17 10.5   &  0.0353     &    NLS1         &   5.06   &   2.79    &  20.30 [20.00 -- 20.48]        & Winter et al. 2009\\
IGR J09253+6929           & 09 25 47.56   & $+$69 27 53.6   &  0.039       &    Sy1.5           &   $<$1.94   &   0.05   &  23.15 [22.58 -- 23.63]   & Malizia et al. 2011\\
IGR J09446-2636            & 09 44 37.02   & $-$26 33 55.4    &  0.1425     &    Sy1.5           &    $<$2.34  &    0.5    & {\bf 20.81}                       & Malizia et al. 2011 \\
NGC 2992                         & 09 45 42.05   & $-$14 19 35.0    &  0.0077     &    Sy2    &   6.51   &   1.20$^{\star}$   &  21.90 [21.88 -- 21.92]   & Yaqoob et al. 2007 \\
\hline
\end{tabular}
\end{sideways}
\end{table} 

\clearpage

\begin{table}
%\rotcaption[]{continued}
\centering
\begin{sideways}
 \begin{tabular}{l l l  c l l l c l }
  \hline\hline
 Name              &      RA     & Dec          &  $z$  &   Class   & F$_{H}^{\dagger}$  & F$_{S}^{\dagger\dagger}$ & Log N$_H^{\diamond}$  & Reference\\
\hline
LEDA 170194                 & 12 39 06.32   & $-$16 10 47.8    &  0.0367       &    Sy2                  &   3.77   &   2.00   &  22.46 [22.41 -- 22.62]    & De Rosa et al. 2008  \\      
NGC 4593                       & 12 39 39.43   & $-$05 20 39.3    &  0.009         &    Sy1                 &   7.12   &   3.72   &  {\bf 20.30}                      &   Molina et al. 2009  \\
IGR J12415-5750          & 12 41 25.74   & $-$57 50 03.5    &  0.0242        &    Sy1.5               &   2.04   &   0.77  &  {\bf 21.48}                       &   Molina et al. 2009\\
IGR J1248.2-5828         & 12 47 57.84   & $-$58 30 00.2    &  0.028       &    Sy1.9                  &   1.02   &   0.40  &  22.35  [22.23 -- 2.48]     &   Landi et al. 2010a   \\
NGC 4748                        & 12 52 12.40   & $-$13 24 53.0    &  0.0146     &    NLS1                &   1.27   &   0.34  &  {\bf 20.56}                       &  Panessa et al. 2011 \\
ESO 323-32                    & 12 53 20.19   & $-$41 38 07.5    &  0.016        &    Sy2                   &  1.81    &   0.40   &    25.00 [\it fixed]            &   Comastri et al. 2010 \\
3C 279                             & 12 56 11.17   & $-$05 47 21.5    &  0.5362      &    QSO/Blazar     &   2.15    &    0.60$^{\star}$    & {\bf 20.35} &  Donato et al. 2005  \\
\hline
\end{tabular}
\end{sideways}
\end{table} 

\clearpage

\begin{table}
%\rotcaption[]{continued}
\centering
\begin{sideways}
 \begin{tabular}{l l l  c l l l c l }
  \hline\hline
 Name              &      RA     & Dec          &  $z$  &   Class   & F$_{H}^{\dagger}$  & F$_{S}^{\dagger\dagger}$ & Log N$_H^{\diamond}$  & Reference\\
\hline
IGR J14471-6319           & 14 47 14.88   & $-$63 17 19.2    &  0.038          &    Sy2              &   1.36     &   0.39      &  22.39  [21.83 -- 22.65]      &  Malizia et al. 2007    \\      
IGR J14515-5542           & 14 51 33.13   & $-$55 40 38.4     &  0.018     &    Sy2                  &   2.08       &   0.53    &  21.52  [21.40 -- 21.70]      &    De Rosa  et al. 2011      \\
IGR J14552-5133           & 14 55 17.80   & $-$51 34 17.0     &  0.016     &    NLS1               &   1.45       &   0.96    &  {\bf 21.53}                         &  Panessa et al. 2011  \\      
IGR J14561-3738           & 14 56 08.43   & $-$37 38 52.4     &  0.024     &    Sy2                  &   1.55       &   0.002  &  $>$24                               &   Sazonov et al. 2008    \\
IC 4518A                          & 14 57 41.16   & $-$43 07 55.2     &  0.0163   &    Sy2                 &   2.08       &   0.29    &  23.15 [23.11 -- 23.23]       &   De Rosa et al. 2008  \\      
MKN 841                          & 15 04 01.20   & $+$10 26 16.1    &  0.0364   &    Sy1.5              &   $<$3.40 &   1.62       &  {\bf 20.84}                      &  Piconcelli et al. 2005  \\      
IRAS 15091-2107          & 15 11 59.80   & $-$21 19 02.0    &  0.0446   &    NLS1                  &   2.44       &   0.90       &  21.15 [21.00 -- 21.25]    &   Panessa et al. 2011\\
\hline
\end{tabular}
\end{sideways}
\end{table}

\clearpage

\begin{table}
%\rotcaption[]{continued}
\centering
\begin{sideways}
 \begin{tabular}{l l l  c l l l c l }
  \hline\hline
 Name              &      RA     & Dec          &  $z$  &   Class   & F$_{H}^{\dagger}$  & F$_{S}^{\dagger\dagger}$ & Log N$_H^{\diamond}$  & Reference\\
\hline
IGR J18259-0706$^{(f)}$  & 18 25 57.58   & $-$07 10 22.8    &  0.037     &    Sy1                 &   1.36       &   0.54        &  21.91 [21.87 -- 21.93]           & this work\\      
IGR J18311-3337           & 18 31 14.75   & $-$33 36 08.5    &  0.0687    &    Sy2                 &  1.66         &   0.026    & 22.15 [21.60 -- 22.81]        &  Malizia et al. 2011   \\
PKS 1830-211                  & 18 33 39.92   & $-$21 03 39.9    &  2.507     &    QSO/Blazar   &   4.83       & 1.00         &  22.29 [22.23 -- 22.35]        &     De Rosa et al. 2005 \\
3C382                                & 18 35 03.39   & $+$32 41 46.8   &  0.0579   &    Sy1                 &   5.91       &   5.98      &  {\bf 20.79}                          & Grandi et al. 2006   \\
ESO 103-35                      & 18 38 20.30   & $-$65 25 41.0    &  0.0133   &    Sy2                 &   8.15       &   2.40      &  23.30 [23.25 -- 23.29]         & De Rosa  et al. 2011\\
3C 390.3                            & 18 42 08.99   & $+$79 46 17.1   &  0.0561   &    Sy1.5                 &   6.13    &   2.14      &  {\bf 20.63}                           &   Molina et al. 2009 \\
{\bf ESO 140-43}               & 18 44 54.01  & $-$62 21 53.2     &  0.0142   &   Sy1.5                  &  3.93      &   9.50     & 23.04 [22.70 -- 23.15]          & Ricci et al. 2010\\
\hline
\end{tabular}
\end{sideways}
\end{table}

\clearpage

\begin{table}
%\rotcaption[]{continued}
\centering
\begin{sideways}
 \begin{tabular}{l l l  c l l l c l }
  \hline\hline
 Name              &      RA     & Dec          &  $z$  &   Class   & F$_{H}^{\dagger}$  & F$_{S}^{\dagger\dagger}$ & Log N$_H^{\diamond}$  & Reference\\
\hline
NGC 7172                        & 22 02 01.90   & $-$31 52 11.6   &  0.0087     &    Sy2                        &   7.93           &  4.30   &  22.86 [ 22.85 -- 22.87]     & De Rosa et al. 2012 \\      
BL Lac                               & 22 02 43.29   & $+$42 16 40.0   &  0.686      &    BL Lac                  &   2.40          &   2.02    & {\bf 21.44}                        & Donato et al. 2005             \\
IGR J22292+6647           & 22 29 13.84   & $+$66 46 51.5   &  0.112      &    Sy1.5                      &   1.11           &   0.69   &  {\bf 21.68}                       &  Landi et al. 2010a   \\
NGC 7314                         & 22 35 46.23   & $-$26 03 00.9    &  0.0048    &    Sy1.9                     &   3.02        &   1.91      &  22.02 [21.95 -- 22.07]     & Dewangan et al. 2005\\      
{\bf IGR J22367-1231}    & 22 36 46.49   &  $-$12 32 42.6    & 0.0241    & Sy1.9                          &  2.76         &     1.09    &  22.42 [22.32 -- 22.53]     &  Malizia et al. 2011  \\
3C 452                               & 22 45 49.10   & $+$39 41 15.0   &  0.0811     &    Sy2                       &   3.00        &   0.50      &  23.77  [23.70 -- 23.83]    & Isobe et al. 2002  \\      
IGR J22517+2218           & 22 51 53.50   & $+$22 17 37.3   &  3.668       &    QSO/Blazar       &  3.15     &   0.26              &  22.48 [22.00 -- 22.70]        & Bassani et al. 2007  \\
\hline
\end{tabular}
\end{sideways}
\end{table}

\end{document}